\documentclass[aps,prl,twocolumn,groupedaddress]{revtex4-1}

\usepackage{amssymb,amsmath,amsthm}
\usepackage{bm}

\begin{document}


\title{New bounds on neutrino electric millicharge from limits on neutrino magnetic moment}


\author{Alexander I. Studenikin}
\email[]{studenik@srd.sinp.msu.ru}
\affiliation{Department of Theoretical Physics, Faculty of Physics, Moscow State University, Moscow 119991, Russia}
\affiliation{Joint Institute for Nuclear Research, Dubna 141980, Moscow Region, Russia}


\date{\today}

\begin{abstract}

Using the new limit on the
neutrino anomalous magnetic moment recently obtained by the GEMMA
experiment on measurements of the  cross section for the reactor antineutrino
scattering on free electrons, we get a
new direct upper bound on the neutrino
millicharge $\mid q_{\nu} \mid < 1.5 \times 10^{-12} e_0$.
This is a factor of 2 more stringent constraint than the previous bound obtained from the TEXONO
reactor experiment data that is included to the Review of Particle Properties 2012.
We predict that with data from the ongoing new phase of the GEMMA experiment
the upper bound
 on the neutrino millicharge will be reduced to $\mid q_{\nu} \mid
< 3.7 \times 10^{-13}e_0$ within two years.
We also predict that with the next phase of the considered experiment
the upper bound on the millicharge will be reduced
by an order of magnitude over the present bound and reach the level $\mid q_{\nu} \mid < 1.8 \times 10^{-13}e_0$ within approximately four years.

\end{abstract}

\pacs{14.60.St}

\maketitle

\section{Introduction}
The discovery of the Higgs boson provides an convincing experimental
confirmation of the solid status of the Standard Model. Therefore,
now one can consider the neutrino to be the only particle
that really exhibits properties beyond the Standard Model. In
addition to the experimentally confirmed nonzero mass, flavour
mixing and oscillations the neutrino nontrivial electromagnetic
properties, once confirmed, would provide a clear indication for
physics beyond the Standard Model.

 Within the
Standard Model neutrinos are massless and have ``zeroth"
electromagnetic properties. However, it is well known that in
different extensions of the Standard Model a massive neutrino has
non trivial electromagnetic properties (for a review of the
neutrino electromagnetic properties see
\cite{Giunti:2008ve,Broggini:2012df}). That is why it is often
claimed that neutrino electromagnetic properties open ``a window
to new physics" \cite{Studenikin:2008bd}.

The neutrino electromagnetic interactions, in addition of being a
powerful tool in exploring beyond the Standard Model frontier, can
generate important effects when neutrinos propagate for long
distances in presence of magnetic fields and media. Therefore,
there are two main approaches for studying the neutrino
electromagnetic properties. The first approach is based on
consideration possible electromagnetic contributions to neutrino
processes in extreme astrophysical environments. A detailed
discussion of the astrophysical option of constraining neutrino
electromagnetic properties can be found in \cite{Raffelt:1996wa}.

The second approach assumes high precision measurements of
neutrino interaction cross section in the terrestrial laboratory
experiments in which the electromagnetic interaction contributions
are hoped to be observed in addition to the main contributions due
to weak interactions. A review on the relevant present results on
the upper bounds on the neutrino effective magnetic moment can be
found in \cite{Giunti:2008ve,Broggini:2012df}.

Note that there are several attempts in the literature aimed to
investigate new promising possibilities for providing more
stringent constraints on the neutrino electromagnetic properties
based on the existed experimental data on ${\bar {\nu}}-e$
scattering. For instance, an interesting possibility for getting
more stringent bound on the neutrino magnetic moment from ${\bar
{\nu}}-e$ scattering experiments based on the ``dynamical zeros"
appear in the Standard Model scattering cross section was
discussed in \cite{Bernabeu:2004ay}. Another possibility was
discussed in \cite{Wong:2010pb} where it was claimed that electron
binding in atoms (the ``atomic ionization'' effect in neutrino
interactions on Ge target) can significantly increase the
electromagnetic contribution to the differential cross section
with respect to the free electron approximation. However, detailed
considerations of the atomic ionization effect in (anti)neutrino
atomic electron scattering experiments presented in
\cite{Voloshin:2010vm,Kouzakov:2010tx,Kouzakov:2011ig,Kouzakov:2011ka,Kouzakov:2011vx}
show that the effect is by far too small to have measurable
consequences even in the case of the low energy threshold of 2.8
keV reached in the GEMMA experiment \cite{Beda:2012zz}.

In this letter we discuss the possibility to obtain constraints on the neutrino
electromagnetic properties using the recent data as well as
the expected results from the GEMMA collaboration on measurements of the reactor antineutrino
scattering off electrons. The goal of the GEMMA experiment is to constrain from above (or discover) the neutrino anomalous magnetic moment $\mu^{a}_{\nu}$. Using the
recently reported \cite{Beda:2012zz} by the GEMMA collaboration
bound on the neutrino magnetic moment
we derive a new direct bound on the neutrino millicharge absolute value
$\mid q _{\nu}\mid < 1.5 \times 10^{-12
} e_0 $, where $e_0$ is the
absolute value of the electron charge. This is a factor of 2 more stringent constraint than the previous bound \cite{Gninenko:2006fi} obtained from the TEXONO reactor experiment data \cite{Li:2002pn} and included by the  Particle Data Group Collaboration to the Review of Particle Physics \cite{Beringer:1900zz}.

As a matter of fact, the scale on which the millicharge is probed within the used scheme depends on the attained scale of the magnetic moment $\mu^{a}_{\nu}$ and the electron recoil energy threshold of the experiment. With the expected future progress of the new ongoing phase of the GEMMA experiment \cite{Beda:2013mta} we predict that in two years the neutrino electric charge will be bound on the level of
$\mid q _{\nu}\mid < 3.7 \times 10^{-13
} e_0 $, and even on the level $\mid q _{\nu}\mid < 1.8 \times 10^{-13
} e_0 $ with the next phase of this experiment within approximately four years from now.

Note that the possibility to constrain neutrino charge from the results
of experiments searching for neutrino magnetic moment was considered in \cite{Babu:1993yh}.
It has been shown, using results of the Big European Bubble Chamber beam dump experiment,
that from a consideration of the elastic scattering $\nu_{\tau} e^{-}\rightarrow \nu_{\tau} e^{-}$
the tau-neutrino electric charge may be bound by $\mid q _{\nu_{\tau}}\mid < 4 \times 10^{-4
} e_0 $. In that case, contrary to the case consider below in this paper, the weak
contributions to the cross section are too small and can be ignored.
A direct experimental limit on the electric charge of the electron
antineutrino  $\mid q _{\nu}\mid < 3.7 \times 10^{-12
} e_0 $ has been obtained as a by product result in \cite{Gninenko:2006fi} where
constraints on millicharged hypothetical particles from the TEXONO reactor experiment were derived.
This limit is presently included by the  Particle Data Group Collaboration to the Review of Particle Physics \cite{Beringer:1900zz}.

The discussed above constraints should be compared with those obtained
from direct accelerator
searches, charged leptons anomalous magnetic moments, stellar
astrophysics and primordial nucleosynthesis (some of that can be in general less
stringent) \cite{Davidson:1991si, Babu:1992sw}:
\begin{equation} q_{\nu}\leq  10^{-6} - 10^{-17} e_0.
\end{equation}
The most severe indirect constraints on the electric charge of
the neutrino
\begin{equation} q_{\nu}\leq 10^{-21} e_0,
\end{equation}
are obtained assuming electric charge conservation in neutron beta
decay $n\rightarrow p+e^-+\nu_{e}$,  from the neutrality of matter
(from the measurements of the total charge $q_{p}+q_{e}$)
\cite{Marinelli:1983nd} and from the neutrality of the neutron
itself \cite{Baumann:1988ue}. A detailed discussion of different constraints on the neutrino
electric charge can be found in
\cite{Raffelt:1996wa,Raffelt:1999gv}.

\section{Electrically millicharged neutrino}

It is usually believed  that the neutrino has a zero electric charge. This can be attributed to gauge invariance and anomaly cancelation constraints imposed in the Standard Model. However, if the neutrino has a mass, the statement that the neutrino electric charge is zero is not so evident as it meets the eye. In theoretical models with the absence of hypercharge quantization the electric charge also gets ``dequantized'' and as a result neutrinos may become electrically millicharged particles. A detailed discussion of theoretical models predicted the millicharged neutrinos as well as possible experimental aspects of this problem can be found in many papers \cite{Bernstein:1963qh}-\cite{Holdom:1985ag}. See also \cite{Giunti:2008ve} for a review on this topic.

\section{Bound on neutrino millicharge from the GEMMA experiment}

Consider a massive neutrino with non-zero electric millicharge
$q_{\nu}$ that induces an additional electromagnetic interaction
of the neutrino with other particles of the Standard Model. Such a
neutrino behaves as an electrically charged particle with the
direct neutrino-photon interactions, additional to one produced by
possible neutrino non-zero (anomalous) magnetic moment $\mu_{\nu}$ that is
usually attributed to a massive neutrino.

If there is no special mechanism of ``screening" of these new
electromagnetic interactions then the neutrino will get a normal
magnetic moment predicted within the Dirac theory of an
electrically charged spin-$\frac{1}{2}$ particle
\begin{equation}\label{mu_q}
\mu_{\nu}^{q}=\frac{q_{\nu}}{2m_{\nu}}
\end{equation}
that is proportional to the neutrino millicharge $q_{\nu}$, here
$m_{\nu}$ is the neutrino mass. In general, this new contribution to the
neutrino magnetic moment should be added to the neutrino anomalous
magnetic moment $\mu_{\nu}^{a}$ that can be generated by the
vacuum polarization loop interactions within different theoretical
models beyond the Standard Model. We recall here that in the
initial formulation of the Standard Model a neutrino is the
massless particle and its magnetic moment is zero. Within the
easiest generalization of the $SU(2)_L \times U(1)_Y$ Standard
Model for a massive neutrino the contribution to the anomalous
magnetic moment is produced by the $\nu - W - e$ loop diagramme.

Thus, for a millicharged massive neutrino one can expect that the
magnetic moment contains two terms,
\begin{equation}\label{mu_q_mu_a}
\mu_{\nu}^{}=\mu_{\nu}^{q} + \mu_{\nu}^{a},
\end{equation}
where in the case of the Dirac neutrino \cite{Fujikawa:1980yx}
\begin{equation}\label{m_e_mom_i_j}
 \mu^{a}_{\nu} = \mu^{D}_{\nu}=\frac{e_0 G_F m_{\nu}}{8\sqrt {2} \pi ^2}
\approx 3.2\times 10^{-19}
  \Big(\frac{m_\nu}{1 \ eV}\Big) \mu_{B}
\end{equation}
is a tiny value for any reasonable scale of $m_\nu$ consistent
with the present neutrino mass limits ($\mu_{B}=\frac{e_0}{2m_e}$
is the Bohr magneton).

Here we recall that there is a hope of both theorists and
experimentalists that new interactions beyond the Standard Model
might reasonably increase the anomalous part of the neutrino
magnetic moment to the level that could be checked by new
terrestrial laboratory experiments in the near future.

Now we consider the direct constraints on
the neutrino millicharge obtained using data on the neutrino electromagnetic
cross section in the GEMMA experiment.
 It is important to note that although in the case of a millicharged
neutrino two terms, i.e. normal and anomalous magnetic moments,
 sum up in the total expression
(\ref{mu_q_mu_a}) for the magnetic moment, however these two contributions
should be treated separately when one considers the electromagnetic
contribution to the   scattering cross section.
The point is that the normal magnetic moment contribution is accounted for
automatically when one considers the direct neutrino millicharge to the
electron charge interaction.

The prescription to obtain the bound on the neutrino millicharge from the
experimental data on the ${\bar{\nu}}-e$ cross section is as follows.
One first compares the magnetic moment cross section
$\left(\frac{d\sigma}{dT}\right)_{\mu^{a}_{\nu}}$ with the
Standard Model weak contribution to the cross section
$\left(\frac{d\sigma}{dT}\right)_{weak}$. From the fact that
the   experimental data on the cross section,
for the presently achieved electron recoil energy threshold $T$,
shows no deviation from the predictions of the Standard Model a limit on the neutrino magnetic
moment is obtained. Then one should compare the magnetic moment contribution
to the cross section, $\left(\frac{d\sigma}{dT}\right)_{\mu^{a}_{\nu}}$, and the contribution due to the neutrino millicharge, $\left(\frac{d\sigma}{dT}\right)_{q_{\nu}}$, and account that the later is also not visible at
the present experiment. In order not to contradict to the experimental data
the cross section $\left(\frac{d\sigma}{dT}\right)_{q_{\nu}}$
should not accede the cross section $\left(\frac{d\sigma}{dT}\right)_{\mu^{a}_{\nu}}$ anyway.
Thus, the obtained upper limit on the neutrino millicharge depends on the achieved upper limit
on the neutrino (anomalous) magnetic moment and the electron recoil energy threshold of
the ${\bar{\nu}}-e$ experiment.

Consider the latest results \cite{Beda:2012zz} of the GEMMA collaboration on
the neutrino magnetic moment. Within the presently reached electron recoil energy threshold of
\begin{equation}\label{1_T}
T \sim 2.8 \ keV
\end{equation}
the neutrino magnetic moment is bounded from above by the value
\begin{equation}\label{mu_bound}
\mu_{\nu}^{a} < 2.9 \times 10^{-11} \mu_{B}. \end{equation}.
In order to get from these data the limit on
the neutrino millicharge we compare the mentioned above two cross sections, $\left(\frac{d\sigma}{dT}\right)_{\mu^{a}_{\nu}}$ and $\left(\frac{d\sigma}{dT}\right)_{q_{\nu}}$.
The expression for the neutrino magnetic moment cross section can be found in \cite{Vogel:1989iv},
for our present needs only the term proportional to $\frac{1}{T}$ matters,
\begin{equation}\label{sigma_mu_1_T}
\left(\frac{d\sigma}{dT}\right)_{\mu^{a}_{\nu}} \approx
\pi\alpha^{2}\frac{1}{m_{e}^{2}T}
\left(\frac{\mu^{a}_{\nu}}{\mu_{B}}\right)^{2},
\end{equation}
here $\alpha $ is the fine structure constant. For the corresponding neutrino millicharge-to-charge
cross section we obtain (see also  \cite{Berestetskii:1979aa})
\begin{equation}\label{sigma_q_e}
\left(\frac{d\sigma}{dT}\right)_{q_{\nu}}\approx 2\pi\alpha
\frac{1}{m_{e}T^2}q_{\nu}^2.
\end{equation}
For the ratio $R$ of the mentioned above cross sections (\ref{sigma_q_e}) and (\ref{sigma_mu_1_T})
we have
\begin{equation}\label{R}
R=\frac{\left(\frac{d\sigma}{dT}\right)_{q_{\nu}}}
{\left(\frac{d\sigma}{dT}\right)_{\mu^{a}_{\nu}}}=
\frac{2 m_e}{T}\frac{\left(\frac{{q}_{\nu}}{e_0}\right)^{2}}
{\left(\frac{\mu^{a}_{\nu}}{\mu_{B}}\right)^{2}}.
\end{equation}
In case there are no observable deviations from the weak
contribution to the neutrino scattering cross section
it is possible to get
the upper bound for the neutrino millicharge demanding that possible effect due to $q_{\nu}$ does not
exceed one due to the neutrino (anomalous) magnetic moment.
This implies that $R<1$ and from (\ref{R}) we get
\begin{equation}\label{q_limit}
q_{\nu}^{2}<\frac{T}{2m_e}\left(\frac{\mu^{a}_{\nu}}{\mu_{B}}\right)^{2}e_0.
\end{equation}
Thus, from the present GEMMA experiment data (\ref{1_T}) and (\ref{mu_bound})
the upper limit on the neutrino millicharge is
set on the level
\begin{equation}\label{q_2012}
\mid q_{\nu} \mid < 1.5 \times 10^{-12} e_0.
\end{equation}
It is interesting to estimate the range of the neutrino millicharge that can be probed
in a few years with the GEMMA-II experiment that is now in preparation and
is expected to get data in 2015  .
It is planed (for details see in \cite{Beda:2012zz,Beda:2013mta}) that the effective threshold will be reduced to $T=1.5 \ keV$ and the sensitivity to the neutrino anomalous magnetic moment will be at the level
$1\times 10^{-11} \mu_{B}$. Then in case no indications for effects of new physics were observed from
(\ref{q_limit}) we predict that the upper limit on the neutrino millicharge will be
\begin{equation}\label{q_bound_GEMMA_2}
\mid q_{\nu} \mid < 3.7 \times 10^{-13} e_0 .
\end{equation}

Now it is also discussed the perspectives of the GEMMA-III experiment
aimed to reach the threshold $T= 350 \ eV$ and the sensitivity to $\mu_{\nu}$ at the level
$9\times 10^{-12} \mu_{B}$ approximately to the year 2018 {\footnote {Vladimir
Brudanin and Alexander Starostin, private communication.}}.
Then  if again there were no deviations from the Standard Model cross section observed the upper limit to the neutrino millicharge will be
\begin{equation}\label{q_bound_GEMMA_3}
\mid q_{\nu} \mid < 1.8 \times 10^{-13} e_0 .
\end{equation}

The obtained new bounds on the neutrino millicharge are more
stringent than many other bounds previously discussed in
literature  \cite{Davidson:1991si, Babu:1992sw}.

\section{Conclusions}

We consider possibility, provided in various extensions of the
Standard Model, that a neutrino is an electrically millicharged
particle. The corresponding non-standard electromagnetic
interactions of such a neutrino generates the additional
contribution to the neutrino electromagnetic scattering off electrons
 that depends on the milicharge. A new upper limit the neutrino magnetic moment
recently obtained by the GEMMA experiment on measurements of the
reactor antineutrino scattering off electrons allows us to get the
new direct upper bound on the neutrino electric millicharge
$\mid q_{\nu} \mid \sim 1.5 \times 10^{-12} e_0$.
With the expected in 2015 new edition of the GEMMA data (GEMMA II)
the neutrino millicharge will be probed on the level of
$\mid q_{\nu} \mid < 3.7 \times 10^{-13} e_0$. Then in a few more years later with
the expected release of the GEMMA III data the neutrino millicharge will be probed on the level of
$\mid q_{\nu} \mid < 1.8 \times 10^{-13} e_0$.
 The obtained
bounds are determined by (\ref{q_limit}) and are in general independent on the neutrino mass
and  are improved with the progress of the neutrino-electron scattering experiments
that can be reached with decreasing of the attained electron recoil energy threshold.

The obtained bound (\ref{q_2012}) on the neutrino millicharge from the recent
experimental data of the GEMMA collaboration
is a factor of 2 more stringent than the reactor neutrino scattering constraint included by the  Particle Data Group Collaboration to the Review of Particle Physics \cite{Beringer:1900zz} and that was obtained by \cite{Gninenko:2006fi} from the TEXONO reactor experiment data \cite{Li:2002pn}. Accordingly, a new bound on the millicharge that will be obtained within a few years with next release of the GEMMA experiment data will be a factor of 10 more stringent than one from the present GEMMA data.

Note that the prediction to obtain in a few years an order of magnitude improvement (see (\ref{q_bound_GEMMA_3}))
in constraining the neutrino millicharge over the present limit (\ref{q_2012}) will be no doubt attained if
the GEMMA collaboration will reach the declared sensitivity
in probing the neutrino magnetic moment in due time. The bound on the neutrino millicharge
(\ref{q_bound_GEMMA_3}) will be reached irrespectively of whether any deviation from the
Standard Model in the cross section ${\bar {\nu}}-e$ were observed or not.

\section{Acknowledgments}

The author is thankful to  Vladimir
Brudanin, Carlo Giunti, Konstantin Kouzakov, Alexander Starostin and Alexey Ternov
for fruitful discussions. This study has been the Russian Science
Foundation grant N. 14-12-00033.

\bibliography{new_nu_charge_2013}

\end{document}